\documentclass[a4paper,twocolumn,11pt,accepted=2023-07-21]{quantumarticle}
\pdfoutput=1
\usepackage[utf8]{inputenc}
\usepackage[english]{babel}
\usepackage[T1]{fontenc}
\usepackage{amsmath}
\usepackage{hyperref}

\usepackage{tikz}
\usepackage{lipsum}
\usepackage{booktabs}
\usepackage{float}

\usepackage[numbers,sort&compress]{natbib}

\begin{document}

\title{Identifying Pauli spin blockade using deep learning}

\author{Jonas Schuff}
\affiliation{Department of Materials, University of Oxford, Oxford OX1 3PH, United Kingdom}
\orcid{0000-0002-8880-2125}
\email{jonas.schuff@materials.ox.ac.uk}
\author{Dominic T. Lennon}
\orcid{0000-0001-8067-4256}
\affiliation{Department of Materials, University of Oxford, Oxford OX1 3PH, United Kingdom}
\author{Simon Geyer}
\orcid{0000-0002-5609-4128}
\affiliation{Department of Physics, University of Basel, 4056 Basel, Switzerland}
\author{David L. Craig}
\orcid{0009-0009-9179-5163}
\affiliation{Department of Materials, University of Oxford, Oxford OX1 3PH, United Kingdom}
\author{Federico Fedele}
\orcid{0000-0003-4466-5576}
\affiliation{Department of Materials, University of Oxford, Oxford OX1 3PH, United Kingdom}
\author{Florian Vigneau}
\orcid{0000-0002-7557-4493}
\affiliation{Department of Materials, University of Oxford, Oxford OX1 3PH, United Kingdom}
\author{Leon C. Camenzind}
\orcid{0000-0002-2278-1915}
\affiliation{Department of Physics, University of Basel, 4056 Basel, Switzerland}
\author{Andreas V. Kuhlmann}
\orcid{0000-0002-7721-5515}
\affiliation{Department of Physics, University of Basel, 4056 Basel, Switzerland}
\author{G. Andrew D. Briggs}
\orcid{0000-0003-1950-2097}
\affiliation{Department of Materials, University of Oxford, Oxford OX1 3PH, United Kingdom}
\author{Dominik M. Zumb\"uhl}
\orcid{0000-0001-5831-633X}
\affiliation{Department of Physics, University of Basel, 4056 Basel, Switzerland}
\author{Dino Sejdinovic}
\orcid{0000-0001-5547-9213}
\affiliation{School of Computer and Mathematical Sciences \& AIML, University of Adelaide, SA 5005, Australia}
\author{Natalia Ares}
\orcid{0000-0003-2588-6322}
\affiliation{Department of Engineering Science, University of Oxford, Oxford OX1 3PJ, United Kingdom}
\email{natalia.ares@eng.ox.ac.uk}
\maketitle

\begin{abstract}

Pauli spin blockade (PSB) can be employed as a great resource for spin qubit initialisation and readout even at elevated temperatures but it can be difficult to identify. 
We present a machine learning algorithm capable of automatically identifying PSB using charge transport measurements. The scarcity of PSB data is circumvented by training the algorithm with simulated data and by using cross-device validation. We demonstrate our approach on a silicon field-effect transistor device and report an accuracy of 96\% on different test devices, giving evidence that the approach is robust to device variability. Our algorithm, an essential step for realising fully automatic qubit tuning, is expected to be employable across all types of quantum dot devices.

\end{abstract}

Electrostatically defined quantum dots are promising candidates for scalable quantum computation and simulation~\cite{loss1998quantum, vandersypen2017interfacing, hensgens2017quantum}. They can achieve universal quantum computation~\cite{veldhorst2015two} with gates reaching high fidelity~\cite{cerfontaine2020closed, noiri2022fast}. Their properties are attractive for large-scale quantum processors, namely all-electrical control, compact size~\cite{vandersypen2017interfacing, philips2022universal,fedele2021simultaneous}, and potential operating temperatures of above 1K~\cite{petit2020universal,yang2020operation, camenzind2022hole}.

Pauli spin blockade (PSB) is often a crucial requirement for spin qubit initialisation and readout. It allows for spin-to-charge conversion, as spin-conserved tunneling leads to current rectification~\cite{hanson2007spins}. We can rely on PSB even at elevated temperatures~\cite{petit2020universal, petit2020high, yang2020operation, camenzind2022hole}. It is thus essential to reliably and efficiently detect PSB. 
But PSB is elusive; in the few-charges regime it can be found in unexpected gate voltage locations or it might be absent, and in the multi-charge regime it has to be found like the proverbial needle in a haystack. Its detection is challenging even for experienced human experimenters since evidence for PSB is subtle and it relies on several factors, including the details of the confinement potential. Those details are affected by fluctuations in the disorder potential due to fabrication variances and defects within the material. 

To achieve true scalability, we need an automatic method for detecting PSB that can be incorporated as a fundamental building block into a fully automatic qubit tuning algorithm. The scarcity of available data makes reliable automation tough. In addition, PSB data tends to be unbalanced, meaning that there are many more examples of measurements in which PSB is not present than examples evidencing PSB. Measurements exhibiting PSB are therefore rare in an already scarce body of data.
An automatic approach would also allow us to gather sufficient data and insight to reveal the factors that determine the presence of PSB, which can be even more difficult to identify in material systems with strong spin orbit coupling.

We demonstrate how to detect PSB using deep neural networks. Deep neural networks were used in the context of charge state identification, coarse and fine tuning and readout~\cite{darulova2020autonomous, moon2020machine, severin2021cross, baart2016computer, kalantre2019machine, zwolak2020autotuning, nguyen2021deep, zwolak2021ray, vanesbroeck2020quantum, teske2019machine, vandiepen2018automated, botzem2018tuning, craig2021bridging, czischek2021miniaturizing,durrer2020automated, lapointe2020algorithm, matsumoto2021noise}, with some approaches using simulated data to train their algorithms~\cite{zwolak2020autotuning, kalantre2019machine, darulova2021evaluation}.
Unlike these cases, automatic PSB detection required us to make use of 
extremely scarce and unbalanced quantum device data. We developed a physics-inspired simulator and introduced cross-device validation to address this challenge. 

We demonstrate our algorithm in a silicon fin field-effect transistor (FinFET) confining holes~\cite{camenzind2022hole}. We show that we can achieve an accuracy of over 96\% on identifying signs of PSB on unseen devices. The data stems from four silicon FinFET devices with different gate dimensions. We designate training devices, from which we extract training data, and testing devices, from which we extract data to test our algorithm.
We discuss the performance of the algorithm for different types of training data, using simulated training data, measured training data, and a combination of both.

\section{Experiment}\label{experiment}

\begin{figure}[h!]
  \centering
  \includegraphics[width=1\linewidth]{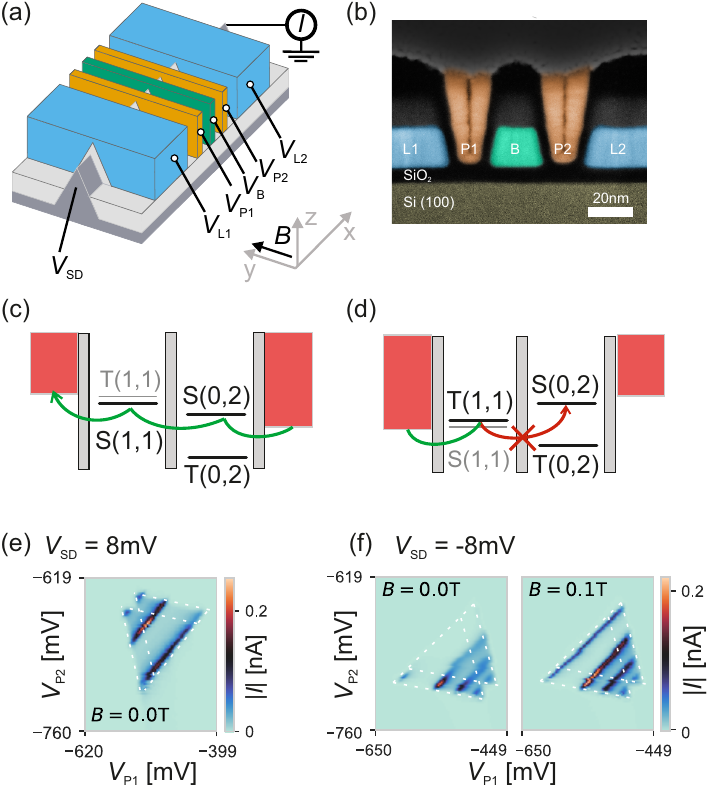}
  \caption{\textbf{Pauli spin blockade in silicon FinFET devices.} \textbf{a} Schematic and \textbf{b} cross-sectional TEM image of a silicon FinFET device. The plunger gates (P1 and P2) accumulate holes in a DQD, the inter-dot barrier is controlled by B and source and drain reservoirs are accumulated using lead gates (L1 and L2). 
  \textbf{c, d} Schematic of a transport cycle in unblocked/blocked configurations respectively. While in the unblocked configuration holes can easily tunnel through the device, spin-conservation blocks the transport through the ground state transition when inverting $V_\mathrm{SD}$ due to the forbidden T$(1,1)-$S(0,2) transition.
  \textbf{e} Bias triangles with positive $V_\mathrm{SD}$. \textbf{f} Bias triangles with negative $V_\mathrm{SD}$. The current at the base line of the triangles, i.e.~current due to the T$(1,1)-$S(0,2) transition, is blocked (left). A finite magnetic field $B=0.1\,$T lifts the blockade (right). We show the absolute value of current for ease of comparison between figures. We outline the bias triangles with white dashed lines to guide the eye.
  } 
  \label{fig:device_and_psb}
\end{figure}

A schematic representation of a silicon FinFET device similar to the ones used in this work and a cross-sectional transmission electron microscope (TEM) image are shown in Figs.\,\ref{fig:device_and_psb}a, b. The devices are fabricated using a CMOS-compatible fabrication process, where a self-alignment technique allows for ultra-small gate length and intrinsically perfect layer-to-layer alignment~\cite{geyer2021self}. 
The fin provides a quasi 1D confinement for holes and a double quantum dot (DQD) can be defined using gate electrodes.
Source and drain reservoirs are formed by lead gates L1 and L2, which accumulate p-type carriers. The plunger gates P1 and P2 allow for control of the hole occupancy.
The inter-dot coupling is controlled by gate B. We perform transport measurements by applying a bias voltage $V_\mathrm{SD}$ between source and drain drawing a current $I$ through the device. Measurements of current as a function of the plunger gate voltages $V_\text{P1}$ and $V_\text{P2}$ are called stability diagrams.
Energetically allowed charge transitions appear as two bias triangles in stability diagrams, see Fig.\,\ref{fig:device_and_psb}e. 
Bias triangles indicate that the device is tuned into the DQD regime.

Four devices with different dimensions (for details see Appendix \ref{app:devices}) were measured at different bias voltages and at temperatures ranging from 20\,mK to 1.5\,K. In all measurements the magnetic field was applied in-plane and perpendicular to the fin as indicated in Fig.\,\ref{fig:device_and_psb}a.

\subsection{Pauli spin blockade}
\label{section_psb}

We look at the $(1,1)-(0,2)$ charge transition, where $(m, n)$ denotes the effective hole occupancy of the left and right dot omitting filled shells. 
For positive $V_\text{SD}$, we expect a hole to tunnel onto the right dot and form a singlet state S(0,2), since the triplet state T(0,2) is energetically unavailable (see Fig.\,\ref{fig:device_and_psb}c). The large energy splitting comes from the fact that the symmetric triplet spin state requires an anti-symmetric orbital state with higher energy. The hole can now tunnel via the S(1,1) state to the left reservoir completing the transport cycle.  
When applying a negative $V_\text{SD}$, we expect a hole to enter the left dot from the reservoir (see Fig.\,\ref{fig:device_and_psb}d). Due to the weak inter-dot coupling S(1,1) and T(1,1) are nearly degenerate, such that both states are accessible. If T(1,1) is occupied, transport through the DQD is blocked since the T$(1,1)-$S(0,2) transition is forbidden by spin conservation and the T(0,2) state is energetically unavailable. 
The effect is also possible in the opposite bias direction, i.e.~a blockade can occur with positive bias voltage for a $(1,1)-(2,0)$ charge transition.

The blockade can be lifted by processes that allow transitions out of the T(1,1) state. In systems dominated by hyperfine interaction~\cite{koppens2005control} or spin-flip cotunneling~\cite{brauns2016anisotropic}, these transitions are enabled at zero magnetic field giving rise to a finite current. For holes in silicon FinFETs, however, spin orbit coupling is the dominant interaction~\cite{geyer2021self, camenzind2022hole}, which lifts the PSB at finite magnetic field due to spin-flip tunneling coupling the T(1,1) states with the S(0,2) state~\cite{danon2009pauli, nadj2010disentangling, li2015pauli}. This mechanism for holes and strong spin orbit coupling in a magnetic field is described in detail in Ref.~\cite{froning2021strong}. Furthermore, independent of the magnetic field, the blockade is lifted when the inter-dot energy level detuning reaches or exceeds the singlet-triplet splitting of the S(0,2) and T(0,2) states.

Signatures of PSB can be observed in Figs.\,\ref{fig:device_and_psb}e, f. Two bias triangles are visible for $V_\text{SD}=8\,$mV. For opposite polarity $V_\text{SD}=-8\,$mV the current at the common base line of the triangles, i.e.~current due to ground state transitions, is strongly suppressed at zero magnetic field. The excited state transitions are visible as parallel stripes away from the common base line. They appear at a detuning exceeding the singlet-triplet splitting and are visible in both bias directions, although the magnitude of the corresponding current might differ due to device asymmetries. The blockade at the base line of the bias triangles is lifted for a magnetic field of $B=0.1$\,T. PSB can be detected by comparing stability diagrams displaying bias triangles at $B=0$ and $B\neq0$ and looking for changes in the base line current. Two stability diagrams showing the same bias triangles at $B=0$ and $B\neq0$ will be called a \textit{pair}. These pairs are the type of data required by our algorithm to identify PSB. A comparison between stability diagrams corresponding to opposite signs of bias voltage could also be used to identify PSB. We expect that this comparison is difficult to use in PSB detection since differences in the transport features might arise from device asymmetries.

\subsection{Simulator}\label{simulator}




A simulator allows us to generate large and diverse data sets needed to train the deep learning algorithm.
Our goal is to simulate pairs of stability diagrams, as introduced in Section \ref{section_psb}.
To achieve this, we first calculate the steady state current~\cite{stoof1996time} with one energy level in each quantum dot. Next, we consider multiple energy levels in each dot and sum the contribution of every possible combination of energy levels to determine the total current.



A simple approach to account for PSB in the simulator involves suppressing the tunneling rate between ground states in each quantum dot. By suppressing those tunneling rates, we identify the ground states with the T(1,1) and S(0,2) configurations. The suppression of the tunnel rates between these two configurations has an analogous effect to that of PSB.
While this can lead to simulated measurements that may be unphysical, 
it enables efficient training of the algorithm. 
We add various sources of noise to the current simulation. We sample each parameter of the simulator, e.g.~lever arms,  from a given range of possible values. We did not perform an optimisation over those sampling ranges to avoid introducing a greater bias to the ouput of the simulator.
A comprehensive description of the simulator is available in Appendix \ref{app:simulator}.




Fig.\,\ref{fig:examples_sim} shows examples of simulated bias triangles. PSB is introduced for $B_\text{sim}=0$ while it is not considered for $B_\text{sim}\neq0$ (Fig.\,\ref{fig:examples_sim}a). 
Each pair of simulated stability diagrams is marked by a grey box in Fig.\,\ref{fig:examples_sim}. Examples of bias triangles where PSB does not occur are shown in Fig.\,\ref{fig:examples_sim}b. 
In this case the difference between measurements in a pair is due to the added noise.

\begin{figure}[h]
\centering
  \includegraphics[width=0.8\linewidth]{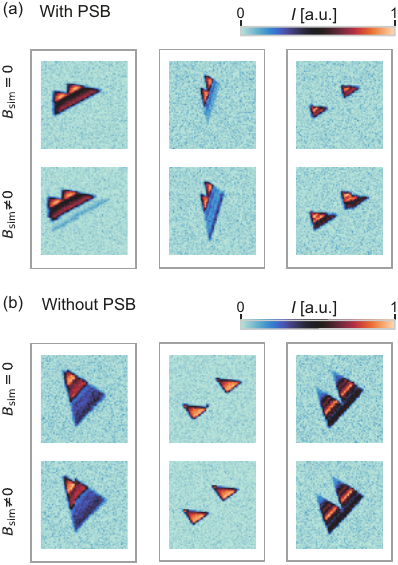}
  \caption{\textbf{Examples of simulated training data.} Each pair in a and b (grey boxes) represents a different set of parameters. The simulated current values are normalized for each of these pairs.  
  \textbf{a} Examples displaying PSB. \textbf{b} Examples in which PSB is not introduced in the simulation. Running the simulator twice with the same parameters results in two similar measurements that only differ due to the noise added to the simulator.
  } 
  \label{fig:examples_sim}
\end{figure}

\subsection{Deep learning}\label{experiment_deeplearning}

The input of the neural network, a deep residual network~\cite{he2016deep} with 18 layers, is a pair consisting of two stability diagrams. Such pairs are defined in Section \ref{section_psb}. The current input values are jointly normalized between 0 and 1 for each pair. 
The neural network outputs a score between 0 and 1. 
A score of 1 (0) corresponds to maximum (minimum) confidence in the occurrence of PSB. The threshold of classification is set at 0.5.

Due to the random sampling of parameters in the simulations and the randomness in the data augmentation process, we expect a high variance in the results of the classification if we train the neural network more than once. We thus train the neural network ten times to obtain ten individual classifiers that we combine into an ensemble of classifiers.
By ensemble classifier we mean a classifier that uses the average score of all individual classifiers as the score of the ensemble. This approach is expected to produce more robust results than individual classifiers. 
Details on the choice of architecture and the training of the neural networks can be found in Appendices~\ref{app:nn_architecure} and \ref{app:deeplearning}.

\section{Results} 





To test our algorithm, we consider a device that is tuned to a double quantum dot regime. Often, the next step is to identify bias triangles which might exhibit PSB. We select gate voltage windows enclosing bias triangles as indicated by rectangles labelled A-C in Fig.~\ref{fig:poc}a. All other bias triangles that could be observed in this device are displayed in Appendix \ref{app:clfresults}. 
To create the pairs introduced in Section \ref{section_psb} we combine the bias triangles delimited by the chosen rectangles with an equivalent version of these measurements at low magnetic field.



 Magnetic field hysteresis can shift the PSB signature away from B=0. This can sometimes occur when using superconducting coils. We thus do not set $B=0$ for these measurements but we choose 9 different equidistant magnetic fields between $B=-0.08\,$T and $B=0.08\,$T. In this way, we create nine pairs of stability diagrams for each gate voltage window A to C. Example pairs are displayed in Fig.\,\ref{fig:poc}b.

\begin{figure}[h!]
\centering
  \includegraphics[width=1\linewidth]{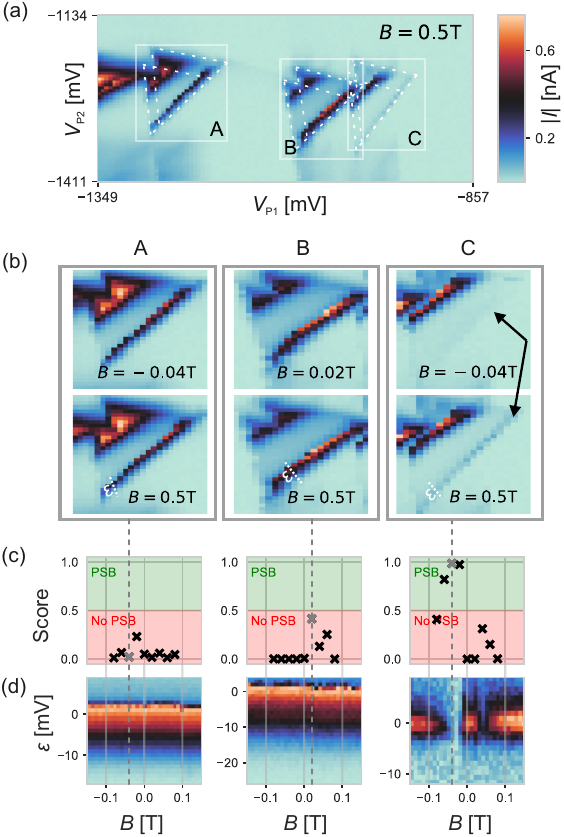}
  \caption{\textbf{Using the classifier to find bias triangles exhibiting PSB.} \textbf{a} Stability diagram taken at $B=0.5$\,T. 
  White rectangles are drawn by a human wishing to classify bias triangles A-C. As in Fig.\,\ref{fig:device_and_psb}, we display the absolute value of current and outline the bias triangles with dashed white lines. 
  \textbf{b} Examples of pairs that are inputs to the classifier. The bottom row shows the bias triangles from Fig.\,\ref{fig:poc}a, which serve as a reference, since we expect PSB to be lifted at this magnetic field value. The same bias triangles are shown in the top row at low magnetic field. Black arrows point at a vanishing common base line indicating the presence of PSB. \textbf{c} Classification of bias triangles A-C. The ensemble of classifiers produces a score for each of the nine pairs corresponding to charge transitions A-C, which are composed by paired measurements at $B=0.5$ T and at values of B close to zero ($B=-0.08$ to 0.08 T). A score of over 0.5 predicts the occurrence of PSB. \textbf{d} Magnetic field dependence of current measurements along the detuning axis, indicated by white dotted lines in Fig.\,\ref{fig:poc}b. A reduction in current confirms the predicted PSB for three pairs of C. We draw a dashed line from the pairs in Fig.\,\ref{fig:poc}b through Fig.\,\ref{fig:poc}c and Fig.\,\ref{fig:poc}d to indicate the magnetic field values corresponding to those pairs. 
  More results can be found in Appendix \ref{app:clfresults}.} 
  \label{fig:poc}
\end{figure}


The predictions obtained by the ensemble of classifiers trained on simulated data as described in Section \ref{experiment_deeplearning} can be seen in Fig.\,\ref{fig:poc}c. All predictions for charge transitions A and B are negative, i.e.~no signs of PSB are detected. For charge transitions in C PSB is detected for pairs with low magnetic field values $B=-0.06,-0.04,-0.02\,$T even though the base line is very faint when PSB is lifted. Our algorithm classifies pairs, not transitions. For charge transition C, some pairs are classified as not exhibiting PSB for those magnetic fields for which the signature of PSB is not apparent. Also, for this transition, the blocked current occurs at $B\neq0$. We assume that the effective magnetic field is 0 for the blocked case and the offset is due to hysteresis.


To confirm the predictions, we measure the current at the base of the bias triangles as a function of the magnetic field and detuning, see Fig.\,\ref{fig:poc}d. The detuning axes, i.e.~the sweep direction of gate voltages $V_\text{P1}$ and $V_\text{P2}$, are indicated as white dotted lines in Fig.\,\ref{fig:poc}b. For charge transitions in C, the current suppression is evident at low magnetic field values. 
This verifies that the corresponding pairs are correctly identified as displaying PSB.
Conversely, no current reduction is observed for bias triangles A and B, confirming the absence of PSB. An automated tuning pipeline could use this information to perform further experiments, such as measurements of Rabi oscillations.
Additional results can be found in Appendix \ref{app:clfresults}.

\subsection*{Benchmarking}\label{generalisation}

We now benchmark the performance of the algorithm for different types of training data, and considering both individual and ensemble classifiers.

We built a data set consisting of 53 pairs, as defined in Section \ref{section_psb}, originating from measurements of four different devices. In this data set, we only included examples that exhibit well shaped bias triangles and measurements that show either clear signatures of PSB or no signatures of PSB at all, so that human experts can verify the correctness of the label. Table \ref{tab:data} shows how many pairs are associated with each device. All pairs used are shown in Fig.\,\ref{fig:all_examples} in Appendix \ref{app:clfresults}.

\begin{table}[htbp]
	\centering
	\begin{tabular}{ccccc}
		\toprule
		  &\multicolumn{4}{c}{Device}  \\ 
		  &  i  &  ii  & iii & iv\\ 
		\cline{2-5}
		 Positive & \multicolumn{1}{|c}{1 }& 15 & 2 & 1\\ 
		    Negative   & \multicolumn{1}{|c}{0 }&16& 14 & 4\\ 
		\bottomrule
	\end{tabular} 
	\caption{\textbf{Structure of experimental data used to benchmark the algorithm.} For each of the devices considered, number of pairs displaying PSB (positive) and not displaying PSB (negative) as assessed by a human judge. Data from device ii was collected over multiple cool-downs. A few individual pairs might show the same charge transitions in different locations in gate voltage space, e.g.~for different tunnel coupling strengths. The data set includes measurements corresponding to $B=-0.04\,$T from Fig.\,\ref{fig:poc} and Fig.\,\ref{fig:more_clf_results}. 
	}
	\label{tab:data}
\end{table}

To study the effect of different training data on the performance of the algorithm, we investigate three cases, which we refer to as Simulated data (\textbf{Sim}), Experimental data (\textbf{Exp}) and Mixed data (\textbf{Mix}). \textbf{Sim} corresponds to the case of training the classifiers with only simulated data. These classifiers are the same as those used for the predictions in Fig.\,\ref{fig:poc}. For \textbf{Exp}, training is performed only with experimental data from the devices listed in Table \ref{tab:data}. \textbf{Mix} is a mix of training with experimental and simulated data; half of the training data is experimental and the other half is simulated.

We augment the training data by random shearing, stretches, crops, contrast and brightness such that there are 50,000 pairs. Details of training and augmentation can be found in Appendix \ref{app:deeplearning}.

Since our data set of measurements is small 
we employ a form of cross-validation in the cases \textbf{Exp} and \textbf{Mix} which we call cross-device validation. This means that each classifier is tested on data from a device that is different from the devices it was trained on. 
Each of these groups of training and testing data forms a \textit{fold}. The process is repeated until all devices have served as a testing device once. Because we don't have both positive and negative pairs from device i we don't use that device as a testing device as computing some specific performance metrics is not possible.
In computation of the cross-validation performance metrics, we weight each fold according to the number of pairs it holds. This can be seen as a form of inverse-variance weighting. In the case \textbf{Sim} we do not need cross-validation since we use only experimental data for testing.

\begin{figure}[ht]
\centering
  \includegraphics[width=1\linewidth]{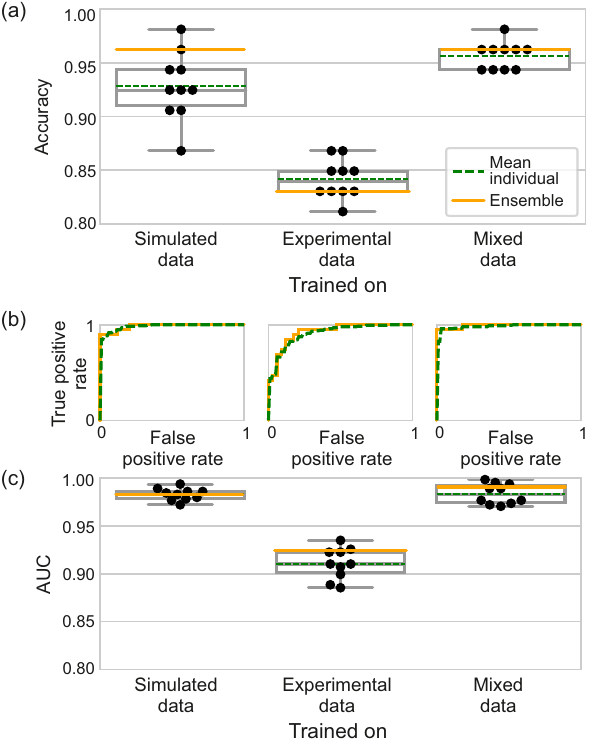}
 \caption{\textbf{Benchmarking results.} \textbf{a} Accuracy. Box plots with  results from individual classifiers plotted as dots, the mean of those is plotted as a dashed green line, and the performance of the ensemble of classifiers as a solid orange line. \textbf{b} Receiver operating characteristic (ROC). For each case, the ensemble ROC is plotted as a solid orange line and the mean individual ROC as a dashed green line. \textbf{c} Area under the curve (AUC). Legend as in Fig.\,\ref{fig:performance}a.} 
  \label{fig:performance}
\end{figure}





We define accuracy as the proportion of correctly classified data in the complete data set. The accuracies of single and ensemble classifiers are shown in Fig.\,\ref{fig:performance}a. Training the classifier with only simulated data (case \textbf{Sim}) leads to an ensemble accuracy of 96.2\%. 
This high accuracy indicates that our simulator produces data that are similar to experimental measurements. The ensemble classifier outperforms the mean of individual classifiers (92.8\%), justifying the use of ensemble classifiers.
Previous classifiers that classify data from quantum devices show accuracies below 90\%\,\cite{darulova2021evaluation}.
The mean accuracy of individual classifiers decreases to 84.2\% for classifiers trained purely on experimental data (case \textbf{Exp}) and the ensemble of classifers achieves an accuracy of only 83.0\%, showing the advantage of a simulator. These lower accuracies indicate over-fitting as a result of small training data sets. The simulated data provides a more representative and diverse data set that prevents this problem. 
Mixing the two data types (case \textbf{Mix}) leads to the same ensemble classifier accuracy as using only simulated data (96.2\%). The mean of the individual classifiers improves to 95.7\% and a lower variance of the individual classifiers is observed.

We find that neural networks trained only on experimental data strongly under-perform those that were trained with only simulated data or a mix of both types of data. 
Thus, the simulated data seems to be the main driver of performance in contrast to the findings in~\cite{darulova2021evaluation}. The superiority of classifiers trained on simulated data may be due to the scarcity of experimental data and specific to the problem of detecting PSB. We train with between 31 and 48 experimental measurements before augmentation in the cases \textbf{Exp} and \textbf{Mix} depending on the fold. In the case \textbf{Sim} (\textbf{Exp}) we use 25,000 (12,500) simulated pairs before augmentation. The influence of the number of pairs used in training is discussed in Appendix~\ref{app:samplesize}.





The accuracy can be affected by the choice of the score threshold so we use other metrics to further analyse our results.
Choosing a score threshold means navigating a trade-off between true positive rate (TPR) and false positive rate (FPR). The receiver operating characteristic (ROC) curve, a plot of TPR against FPR, illustrates this trade-off, see Fig.\,\ref{fig:performance}b. The area under the ROC curve (AUC) is independent of the score threshold and is 1 for a perfect classifier. 
An arbitrary classifier would produce a ROC that is a diagonal with an AUC of 0.5.
In the case \textbf{Sim} we obtain an AUC of 0.983 for the ensemble classifier, see Fig.\,\ref{fig:performance}c. This can be slightly improved by mixing in experimental data (case \textbf{Mix}), leading to an AUC of 0.991. In comparison, only using experimental data (case \textbf{Exp}) gives an ensemble AUC of 0.924. The mean individual classifiers achieve an AUC of 0.983, 0.910, and 0.984 in cases \textbf{Sim}, \textbf{Exp} and \textbf{Mix}, respectively. The results obtained by estimating the AUC are similar to those obtained by calculating the accuracy of the classifiers; training only on experimental data results in under-performing classifiers. 
A comparison with the results obtained using a smaller network can be found in Appendix \ref{app:lenet_results}. An evaluation of performance on simulated data is presented in Appendix \ref{app:test_on_sim}. Classification results for all experimental pairs used here can be found in Fig.\,\ref{fig:all_examples} in Appendix \ref{app:clfresults}.

\section{Discussion}

We train deep neural networks with simulated and experimental data to detect bias triangles that show signs of PSB. We demonstrate that even in the case of extremely limited data, a neural network can be successfully trained to solve this intricate task. 
Cross-device validation allows us to show that the method performs well on unseen devices. 

We find a higher variance of accuracy of individual classifiers when trained on simulated data compared to classifiers trained on real data. This might be due to the limitations of the simulator and could be mitigated by increasing the number of simulations used in training at a larger computational cost. Forming an ensemble prediction 
leads to a high accuracy. 
In contrast to previous work~\cite{darulova2021evaluation}, simulated data seems to be more important for training than experimental data. This might be due to the scarcity of experimental data available for the classification of PSB. The small experimental data set potentially lacks comprehensive information, which could hinder the neural network's ability to learn.

In other types of devices, the signature of PSB might be reversed, i.e.~a maximum leakage current is observed for $B=0$. Our methodology remains applicable to these cases by merely reversing the sequence of the pairs employed in the classification. Competing effects might lead to partial lifts when PSB is present. We can see that a few partial lifts present in the experimental data were correctly classified by our algorithm.

The hurdles in the scarcity of data can be overcome through careful training and high quality simulated data. Scarcity of data could be addressed by the community through open access to data. We expect our algorithm to identify PSB in different types of devices given consistent datasets.

Alternative approaches to the automatic identification of PSB could include both data-driven and non-data-driven methods, such as computer vision techniques or feature engineering. However, due to the limited availability of data and the elusive nature of signatures associated with PSB, these approaches could prove particularly challenging.

The simulator developed in this study holds the potential for various applications, including the development of an algorithm capable of recognizing energy splittings or defining an energy detuning axis. The integration of simple simulators to train machine learning models can be applied to tackle a wide range of quantum device challenges, including feature extraction and efficient measurement of stability diagrams, Rabi chevrons, and EDSR spectroscopy.

In light of the subtlety of the problem of identifying PSB we expect this method to have promising applications in the automation of tuning procedures for spin qubit devices. The classifier could be embedded in a larger tuning algorithm to reliably determine whether a charge transition is promising for PSB and reveal hidden patterns.  

\section*{Acknowledgments}
This work was supported by the Royal Society, EPSRC Platform Grant (EP/R029229/1), the European Research Council (grant agreement 948932), the Swiss NSF Projects 179024 and NCCR SPIN, and the Swiss Nanoscience Institute. The research leading to these results has also received funding from the European Union's Horizon
2020 Research and Innovation Programme \textit{European Microkelvin Platform (EMP)} under Grant Agreement no 824109. J.S.~acknowledges financial support from the EPSRC, Grant Number R72976/CN001. 
We thank the Cleanroom Operations Team of the Binning and Rohrer Nanotechnology Center (BRNC) for their help and support in device fabrication.

We want to thank Edward Laird and Christian Schroeder de Witt for helpful discussions, Brandon Severin for useful discussions about the devices, and Floris Zwanenburg, Matthias Brauns, Rafael Eggli, Mathieu de Kruijf, Sergey Frolov and Azarin Zarassi for sharing experimental data that was used in early stages of this work.

\section*{Data availability}

The data is available at \href{https://doi.org/10.5281/zenodo.7948852}{\url{doi.org/10.5281/zenodo.7948852}}

\section*{Code availability} 
The code used in this publication, including the simulator and the training procedure, is available at \href{https://github.com/oxquantum-repo/identifying-psb}{\url{github.com/oxquantum-repo/identifying-psb}}.

\section*{Author contributions}
J.S., S.G.~and D.T.L.~performed the experiments at the University of Basel under the supervision of L.C.C., A.V.K~and D.M.Z. J.S.~and D.C.~wrote the simulator with guidance from D.T.L., F.F.~and F.V. J.S.~performed the training and evaluation of the neural networks with help from D.T.L.~and under the supervision of D.S., G.A.D.B.~and N.A. The samples were fabricated by S.G.~and A.V.K. The project was conceived by N.A. All authors contributed to the manuscript, and commented and discussed results.



\bibliographystyle{unsrtnat}
\bibliography{reference}

\onecolumn\newpage
 

\twocolumn
\appendix

\section{Device dimensions}\label{app:devices}

\begin{figure}[h]
\centering
  \includegraphics[width=0.8\linewidth]{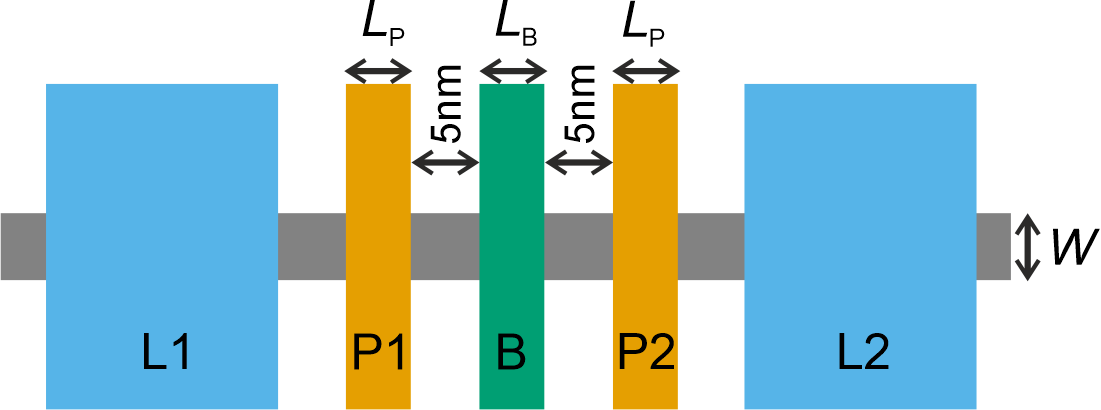}
  \caption{\textbf{Top view schematic of device layout.} Colors and labels are the same as in Figs.~\ref{fig:device_and_psb}a, b.} 
  \label{fig:device_dimensions}
\end{figure}

We use four different silicon FinFET devices with varying dimensions in this work. The dimensions that are varied are the length of the plunger gates $L_\text{P}$, the length of the barrier gate $L_\text{B}$, and the width of the fin $W$. We show a schematic of the top view of the device with the lengths and widths that are varied in Fig.~\ref{fig:device_dimensions}. A top view SEM image of similar devices can be found in \cite{geyer2021self}.
Table \ref{tab:dev_dim} shows the estimated dimensions for each device.

\begin{table}[h]
	\centering
	\begin{tabular}{rcccc}
		\toprule
		  &\multicolumn{4}{c}{Device}  \\ 
		  &  i  &  ii  & iii & iv\\ 
		\cline{2-5}
		 $L_\text{P}$ [nm] & \multicolumn{1}{|c}{20 }& 15 & 15 & 20\\ 
		 $L_\text{B}$ [nm] & \multicolumn{1}{|c}{35 }&25& 35 & 20\\ 
		 $W$ [nm] & \multicolumn{1}{|c}{25}&20& 25 & 10\\ 
		\bottomrule
	\end{tabular} 
	\caption{\textbf{Estimated device dimensions for the different devices.} Illustration of the layout is given in 	Fig.~\ref{fig:device_dimensions}.
	$L_\text{B}$ gives the length of the gap between the plunger gates for device i as it has no barrier gate.}
	\label{tab:dev_dim}
\end{table}

\section{Simulator}\label{app:simulator}


We consider the steady state current through a double quantum dot coupled to fermionic reservoirs. For an energy level $E_A$ in the left dot and an energy level $E_B$ in the right dot we define the detuning as $\epsilon  = E_A - E_B$. These energy levels are seen as the combined charge and spin states of an excess electron. We describe the simulator in terms of electrons but it holds true for holes as in the silicon FinFET used in the main part. We consider an electron to already occupy the right dot. The tunneling rates from the left reservoir (the source) and to the right reservoir (the drain) are $\Gamma_L$ and $\Gamma_R$, and the tunneling rate between the dots is $\Gamma_T$. This situation is shown in Fig.\,\ref{fig:explanation_sim}a. 
\begin{minipage}{\linewidth}
\begin{equation}\label{eq:partialsolution}
    I_\text{partial}=\frac{\Gamma_T^2\Gamma_R}{\Gamma_T^2(2+\Gamma_R/\Gamma_L)+\Gamma_R^2/4+\epsilon^2}.
\end{equation}
\vspace{0mm} 
\begin{figure}[H]
\centering
\includegraphics[width=0.8\linewidth]{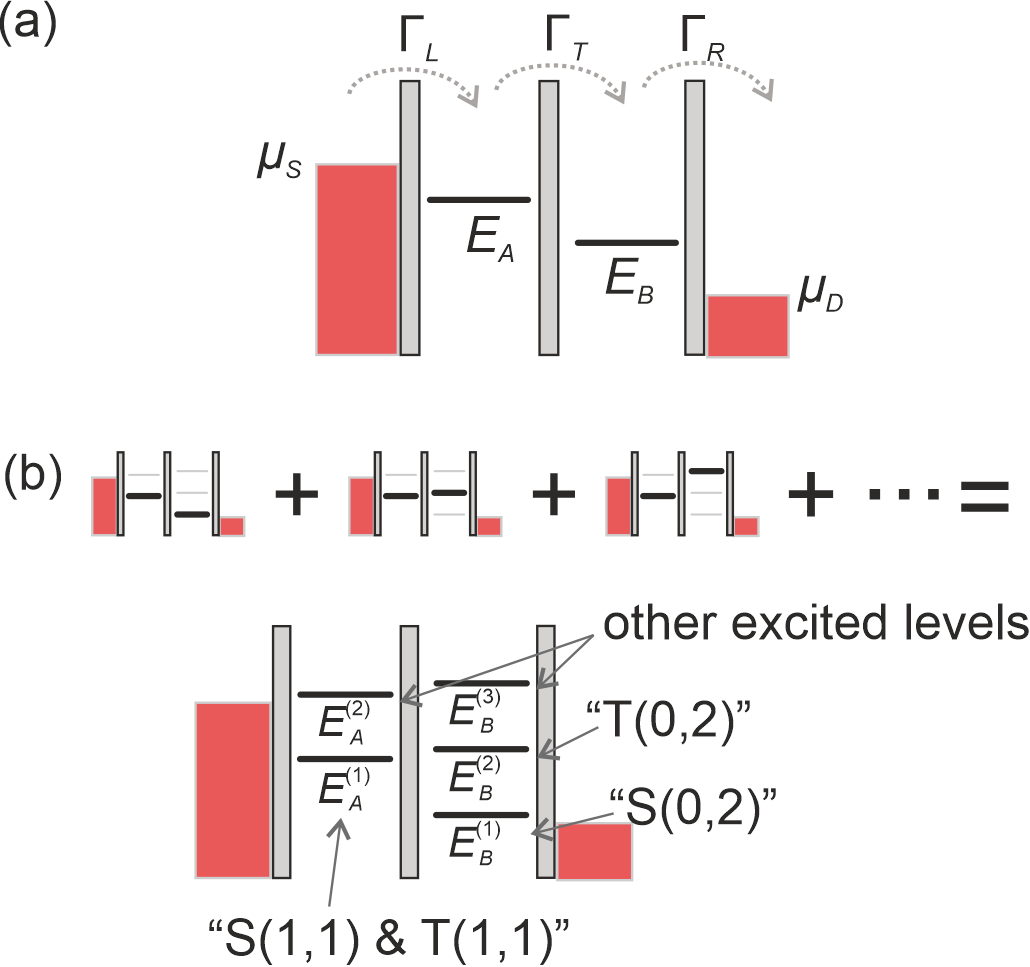}
\caption{\textbf{Explanation of the simulator.} \textbf{a} Fundamental building block. If only one energy level is available in each dot, we can compute the current through a double quantum dot. \textbf{b} Simulation with several levels in each dot. We simply add the contribution of each pair according to \ref{fig:explanation_sim}a. If the DQD is in Pauli spin blockade, lower levels can be identified with singlet and triplet levels, higher levels are excited levels.} 
\label{fig:explanation_sim}
\end{figure}
\vspace{1mm} 
\end{minipage}

The total current consists of contributions due to different energy levels in the left and right dot, as illustrated in Fig.\,\ref{fig:explanation_sim}b. It is given by


\begin{widetext}
\begin{equation}
    I = \sum_{E_A}\sum_{E_B} I_\text{partial}
    =\sum_{i}\sum_{k}\frac{\left(\Gamma_T^{(i,k)}\right)^2\,\Gamma^{(k)}_R}{\left(\Gamma_T^{(i,k)}\right)^2(2+{\Gamma^{(k)}_R}/{\Gamma^{(i)}_L})+{\left(\Gamma^{(k)}_R\right)^2}/{4}+\epsilon_{(i,k)}^2}.
\end{equation}
\end{widetext}

Each level in the left dot $E_A^{(i)}$ is associated with a source tunneling rate $\Gamma_L^{(i)}$. Accordingly, each level in the right dot $E_B^{(k)}$ is associated with a drain tunneling rate $\Gamma_R^{(k)}$ and each pair of energy levels $(E_A^{(i)}, E_B^{(k)})$ is associated with an inter-dot tunneling rate $\Gamma_T^{(i,k)}$. This equation is only used if the lowest energy levels in each dot are in the bias window. Otherwise the DQD is in Coulomb blockade and current will be suppressed by setting $I$ to zero.

Energy levels are computed by considering the gate voltage of each assigned plunger gate $V_{A}, V_B$ and the associated lever arm $L_A, L_B$ that translates voltage to an energy. Additionally, we introduce cross talk terms $C_A, C_B$ which describe the influence of a plunger gate on the other quantum dot, i.e.~the dot that the gate was not intended to be influenced. This way, the ground energy levels $E_A^{(0)}$ and $E_B^{(0)}$ can be computed as 

\begin{equation}
E_A^{(0)}= L_A \cdot V_A + C_B \cdot V_B,
\end{equation}
\begin{equation}
E_B^{(0)}= C_A \cdot V_A + L_B \cdot V_B .
\end{equation}

Each energy level $E_A^{(i)}$ ($E_B^{(k)}$) for $i>0$ ($k>0$) is split from the previous energy level by an energy $E_\text{split}^{(i)}$ ($E_\text{split}^{(k)}$):

\begin{equation}
E_A^{(i)}= E_A^{(i-1)}+ E_\text{split}^{(i)},
\end{equation}
\begin{equation}
E_B^{(k)}= E_B^{(k-1)}+ E_\text{split}^{(k)}.
\end{equation}

We apply white noise $\kappa$ to each energy level with $\kappa \sim \mathcal{N}(1,\,\sigma^{2})$, where $\sigma$ is a sampled parameter. The indices $(i),(k)$ are omitted from now on for readability.

Thermal broadening of the triangles is taken into account by adding the thermal energy $k_\text{B}\mathcal{T}$ to source potential $\tilde{\mu}_S$ and drain potential $\tilde{\mu}_D$ with the Boltzmann constant $k_\text{B}$ and temperature $\mathcal{T}$, $\mu_S=\tilde{\mu}_S+k_\text{B} \mathcal{T}, $ and $ \mu_D=\tilde{\mu}_D+k_\text{B} \mathcal{T}.$ 
The tunnel rates are modified due to the effects of temperature as,

\begin{equation}
\Gamma_L = \tilde{\Gamma}_L f(E_A,\mu_S) ,
\end{equation}
\begin{equation}
\Gamma_R = \tilde{\Gamma}_R [1-f(E_B,\mu_D)] ,
\end{equation}
where $\tilde{\Gamma}_L$ and $\tilde{\Gamma}_R$ are sampled parameters, and $f(\epsilon,\nu)=(1+\exp(\frac{  \epsilon-\nu  }{  k_\text{B} \mathcal{T}  }))^{-1}$ is the Fermi-Dirac distribution. Equally, $\tilde{\Gamma}_T$ is sampled and then rectified to only allow for physically possible transitions in the bias direction:
\begin{equation}\label{eq:rectification}
    \Gamma_T = \begin{cases}
        \tilde{\Gamma}_T, & \text{if }  E_A - E_B \geq 0\\
        0, & \text{else.} 
        \end{cases} 
\end{equation}

Equations \ref{eq:partialsolution} to \ref{eq:rectification} produce one triangle of the pair we need. We can think of this as the cycle $(0,1)\rightarrow(1,1)\rightarrow(0,2)\rightarrow(0,1)$ with $(m,n)$ indicating $m$ electrons (holes) in the left dot and $n$ electrons (holes) in the right dot. There is another cycle possible, namely $(1,2)\rightarrow(1,1)\rightarrow(0,2)\rightarrow(1,2)$. The bias triangle associated with this cycle is shifted due to the electrostatic coupling energy between the dots $E_{C_m}$. This second triangle is simulated by shifting all energy levels by $E_{C_m}$ and repeating the current simulation discussed above. If the bias triangles overlap, only the maximum current is used.

To mimic experimental observations and to create a diverse data set, we get a set of two bias triangles by randomly sampling the scalar parameters $\{L_A, L_B, C_A, C_B, \sigma, \tilde{\mu}_S,\tilde{\mu}_D,\mathcal{T}, E_{C_m}\}$, the vector parameters $\{  \tilde{\mathbf{\Gamma}}_L, \tilde{\mathbf{\Gamma}}_R, \mathbf{E}_\text{split}\}$ and the elements of the matrix $\mathbf{\tilde{\Gamma}}_T$. The dimensions of the vectors and the matrix depend on how many energy levels we consider in each dot, which is also sampled randomly. Overall, the simulator takes between 20 and 50 sampled parameters as input and generates a two dimensional charge stability diagram.

Finally, we add different types of noise to the measurements further to the noise already described. Gaussian blurring is used to smooth the edges of the triangles. The triangles can also move in voltage space due to charge switches or other drift effects. Charge switch noise at random points simulates both effects, see Figure \ref{fig:examples_sim}b bottom left for an example. We also add white noise to the final current values at each simulated point.

To simulate the effect of PSB we add simple rules about where and how much current is allowed. The right lowest level represents the S(0,2) level and we neglect the small splitting of the left lowest energy level into S(1,1) and T(1,1), see Figure \ref{fig:explanation_sim}b, making it only one level. With PSB, tunneling between these two states is prohibited because an electron will eventually occupy the T(1,1) state and will not be able to tunnel to S(0,2). This leads to the rule:
\begin{itemize}
    \item[]For bias triangles in PSB, the tunneling rate between the two lowest energy levels is set to 0.
\end{itemize}
 An electron that is stuck in T(1,1) also blocks all other paths through the double dot. Therefore, a second rule is introduced:
\begin{itemize}
    \item[]We suppress all current at a given point in voltage space if the only available energy levels in the bias window are the lowest ones.
\end{itemize}
 For more details, the code for the simulator is available on our GitHub repository.

\section{Neural Network Architectures}
\label{app:nn_architecure}
We experimented with two neural network architectures: a slightly modified version of ResNet18 and a custom LeNet5-like model. Both models were adapted to accept input images with two channels and to produce outputs for two classes. In this section, we briefly describe the architectures and modifications. For details see our GitHub repository. 

\subsection{Modified ResNet18}

The modified ResNet18 architecture is based on the original ResNet18 model \cite{he2016deep}, with changes to the input and output layers. The first convolutional layer is altered to accept two input channels instead of three, while the final fully connected layer is changed to output two classes instead of 1000. With these changes, the model has 11,174,402 learnable parameters. We use the TorchVision \cite{torchvision} implementation of this model. All results in this paper are based on models with this architecture unless otherwise explicitly stated.

\subsection{Modified LeNet5}
\label{app:lenet_architecture}
The ResNet18 architectures could be considered too large compared to the size of the data set. We therefore consider a much smaller neural network. Our modified 
LeNet5 model \cite{lecun1998gradient} consists of two convolutional layers, each followed by batch normalization, a ReLU activation, and max-pooling. The feature maps are then flattened and passed through three fully connected layers, with ReLU activations after the first and second layers. This leads to 942,500 learnable parameters, which is about an order of magnitude smaller compared to the ResNet18 architecture.

\section{Details of the training procedure}\label{app:deeplearning}

Each training run consists of 50,000 pairs and 100 epochs. We use mini-batches of size 128.


We use the Adam optimizer~\cite{kingma2014adam} with a regularisation factor of 0.001. The optimiser is initialised with a learning rate of 0.001. The learning rate is then decayed by a factor of 0.1 with a scheduler once a plateau in the training loss is reached, with a patience of 5 epochs. We used cross-entropy loss as the objective function, with balanced class weights to account for class imbalance in the training data.

To sample the 50,000 examples for the case where we only use simulated data, we sample 25,000 and augment each image twice. 

When using only experimental data, we augment the available training data (see Table \ref{tab:data}) until we have 50,000 examples. In the case of mixed data, we sample 12,500 examples from the simulator, augment them twice, and then augment the available experimental training data until we have 25,000 examples which gives us 50,000 examples in total. To counteract class imbalance, i.e.~examples with and without PSB, we also weight the classes according to their prevalence in the loss function.

We add random contrast and brightness to all training data and then crop them randomly. Experimental data is additionally randomly sheared and stretched along both axes. 
Every pair is normalised between 0 and 1.

Testing data, i.e.~the corresponding fold in the cross-validation procedure, is not augmented but only normalised.
The neural network and its training is implemented in PyTorch~\cite{pytorch}.

\section{Influence of the number of simulated pairs}\label{app:samplesize}

\begin{figure}
\centering
  \includegraphics[width=0.9\linewidth]{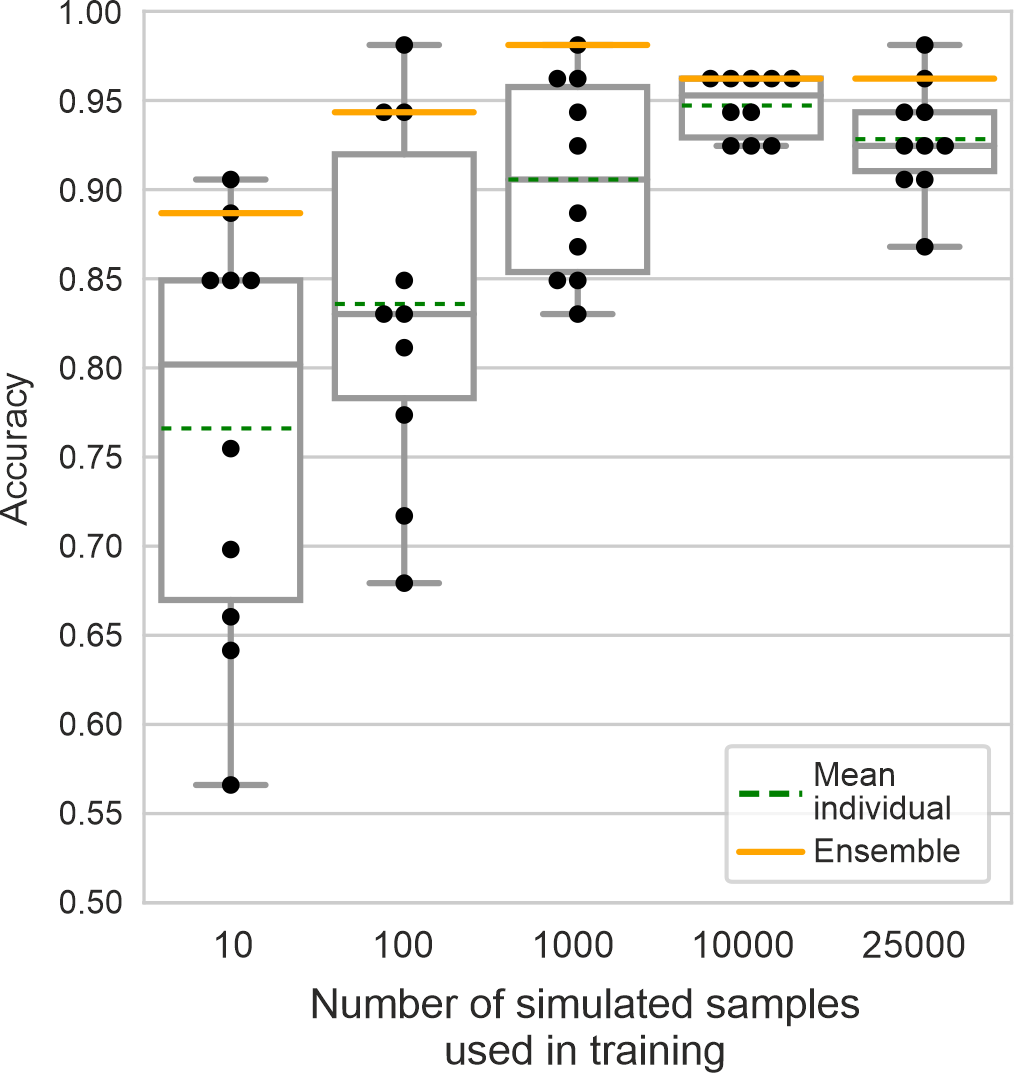}
  \caption{\textbf{Varying simulated sample size.} We show the accuracy for varying sizes of the simulated data set. In all cases the simulated pairs were augmented until 50,000 total pairs are created. 25,000 pairs in the training set corresponds to the analysis shown in the main text in Fig~\ref{fig:performance}.}
  \label{fig:samplesize}
\end{figure}

We investigate the role of the number of simulated pairs when training with only simulated data. We simulate 10, 100, 1,000, or 10,000 pairs and augment them until we have 50,000 pairs. The accuracy for classifiers trained on those data sets in comparison with the sample size used in the main text (25,000 sampled pairs) is shown in Fig.~\ref{fig:samplesize}.

The results show that the main driver of accuracy is the size of the sample set. The number of pairs when training with real data is between 31 and 48, depending on the fold, reaching about 85\% mean accuracy. In comparison, 100 simulated pairs lead to a similar mean accuracy but the spread of accuracies is much larger.

\section{Results from the LeNet5 architectures}
\label{app:lenet_results}

\begin{figure}
\centering
  \includegraphics[width=1\linewidth]{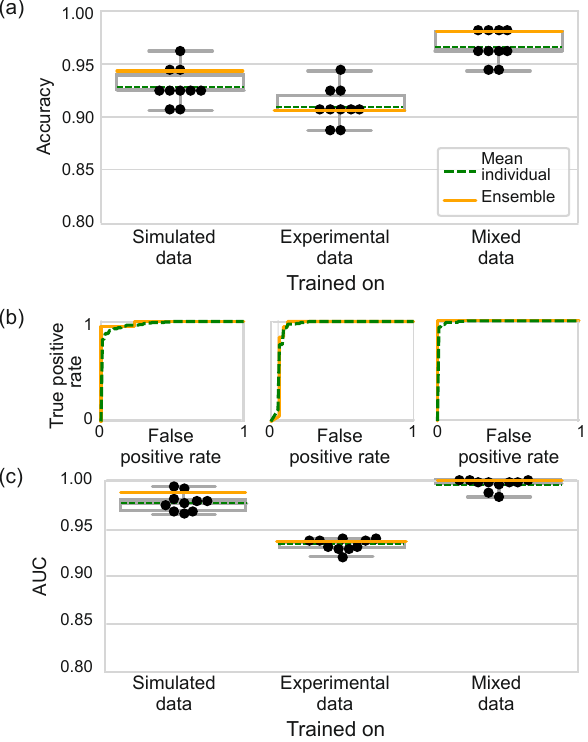}
 \caption{\textbf{Benchmarking results with LeNet architecture.} This is the same analysis as in Fig.~\ref{fig:performance} but with a different neural network architecture. \textbf{a} Accuracy. Box plots with  results from individual classifiers plotted as dots, the mean of those is plotted as a dashed green line, and the performance of the ensemble of classifiers as a solid orange line. \textbf{b} Receiver operating characteristic (ROC). For each case, the ensemble ROC is plotted as a solid orange line and the mean individual ROC as a dashed green line. \textbf{c} Area under the curve (AUC). Legend as in Fig.\,\ref{fig:performance_lenet}a.}
  \label{fig:performance_lenet}
\end{figure}

We repeat the benchmarking as in Section \ref{generalisation} with the smaller LeNet architecture described in Appendix \ref{app:lenet_architecture}. The results are shown in Fig.~\ref{fig:performance_lenet}. The performance in the case \textbf{Exp} is improved significantly, reaching an ensemble accuracy of 90.6\%. This indicates that the size of the ResNet18 architecture leads to over-fitting. However, given the size of our dataset, we are limited in the creation of a holdout dataset to report generalization capability and to tune the hyperparameters of the training without risking overfitting (see Appendix~\ref{app:test_on_sim}).
Still, the ensemble trained in the \textbf{Mix} setting is outperforming the \textbf{Sim} and \textbf{Exp} case, showcasing the usefulness of the simulator. The ensemble classifier of \textbf{Mix} reaches an accuracy of 98.1\% and an AUC of 1.0.

We refrained from optimizing any hyperparameters during the training stage, in an effort to minimize the risk of overfitting. This includes the choice of architecture.

\section{Testing on simulated data}
\label{app:test_on_sim}

To further judge the generalisation performance of the models, we test them on new simulated data. We simulate 1000 pairs of stability diagrams. For \textbf{Exp} and \textbf{Mix} we use models that were trained on all available data, i.e.~we do not separate data into folds.

\begin{figure}
\centering
  \includegraphics[width=1\linewidth]{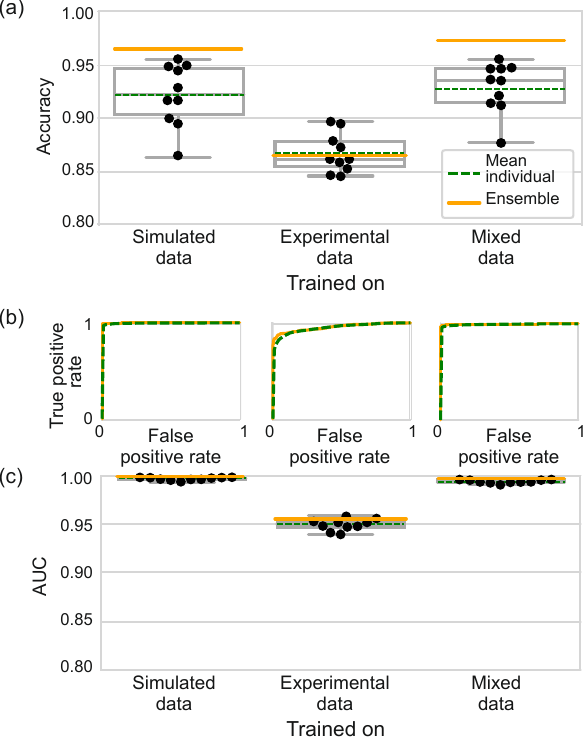}
 \caption{\textbf{Benchmarking results on simulated data with LeNet architecture. }Description as in Fig.~\ref{fig:performance_lenet} but for testing on simulated data.} 
  \label{fig:performance_lenet_on_sim}
\end{figure}

\begin{figure}
\centering
  \includegraphics[width=1\linewidth]{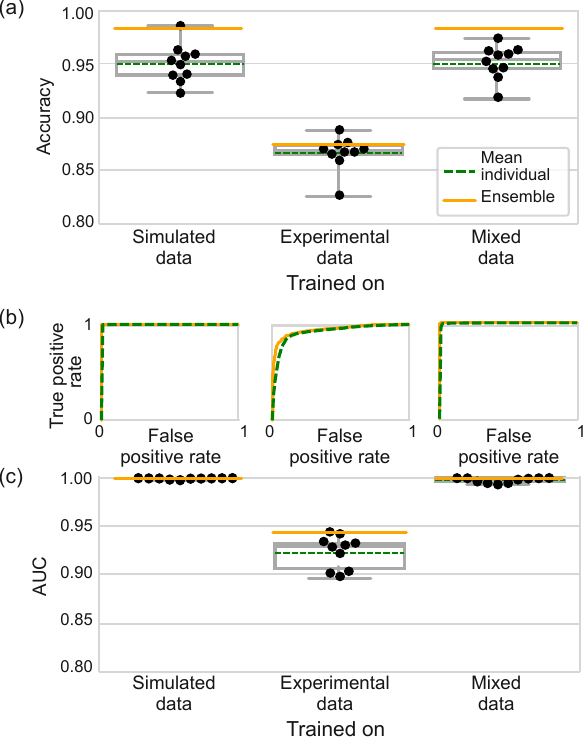}
 \caption{\textbf{Benchmarking results on simulated data with ResNet architecture. }Description as in Fig.~\ref{fig:performance_lenet} but for testing on simulated data.} 
  \label{fig:performance_resnet_on_sim}
\end{figure}

The results are similar to those found testing on experimental data but with notable changes when comparing the models with different sizes. Fig.~\ref{fig:performance_lenet_on_sim} shows the performance of the smaller LeNet models and Fig.~\ref{fig:performance_resnet_on_sim} of the larger ResNet models. With an ensemble accuracy of 86.5\% and an AUC of 0.956, the smaller LeNet model is performing very similarly to the larger ResNet model in the \textbf{Exp} case with an ensemble accuracy of 87.5\% and an AUC of 0.943. For the other cases, the larger ResNet (\textbf{Sim}: ensemble accuracy 98.4\% and AUC 1.0; \textbf{Mix}: ensemble accuracy 98.5\% and AUC 1.0) performs better than the LeNet (\textbf{Sim}: ensemble accuracy 96.5\% and AUC 0.999; \textbf{Mix}: ensemble accuracy 97.3\% and AUC 0.997) which hints at a greater generalisation capability of the larger network when trained with sufficiently diverse data.

\section{More classification results}\label{app:clfresults}

\begin{figure}
\centering
  \includegraphics[width=0.9\linewidth]{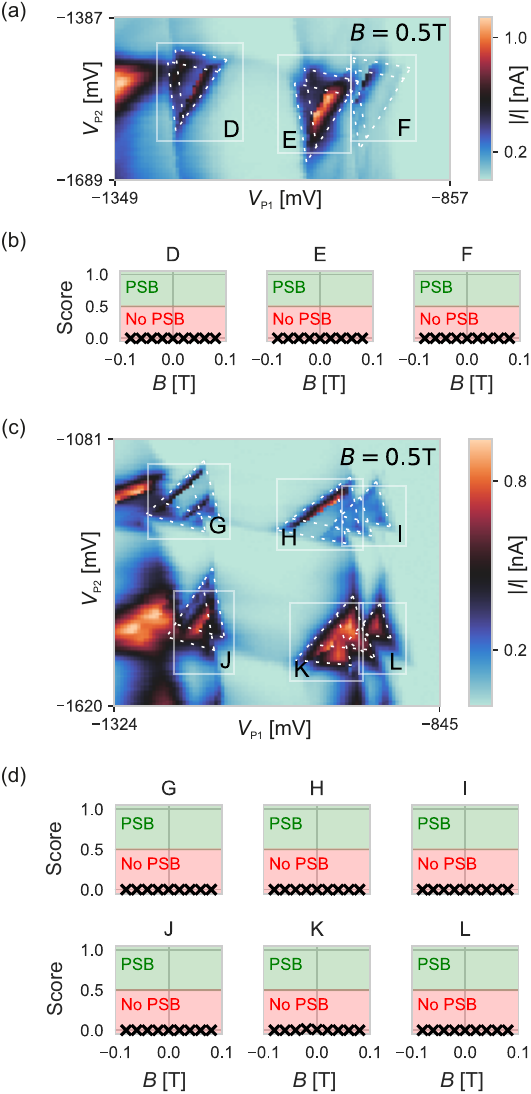}
  \caption{\textbf{More classification results.} \textbf{a} Stability diagram for more bias triangles. \textbf{b} Classification results corresponding to charge transitions in Fig.\,\ref{fig:more_clf_results}a. \textbf{c} Stability diagram for reversed bias. \textbf{d} Classification results in the reversed bias case. As in Figs.\,\ref{fig:device_and_psb} and \ref{fig:poc}, the absolute value of current is shown and dashed white lines outline the bias triangles in Fig.\,\ref{fig:more_clf_results}a and Fig.\,\ref{fig:more_clf_results}c.}
  \label{fig:more_clf_results}
\end{figure}

Fig.\,\ref{fig:more_clf_results} shows more classification results that correspond to section \ref{generalisation} in the main text. Fig.\,\ref{fig:more_clf_results}a shows the stability diagrams with the identified bias triangles as white boxes and Fig.\,\ref{fig:more_clf_results}b shows the predictions of the classifier, which predicted no PSB for charge transitions D-F.

We repeat the same experiment for reversed bias. Fig.\,\ref{fig:more_clf_results}c shows the stability diagram with identified bias triangles. We call them G-L to distinguish them from the measurements in the main text even though they correspond to charge transitions A-F.

Here, all bias triangles were classified as not having PSB, shown in Fig.\,\ref{fig:more_clf_results}d.

\begin{figure*}[ht]
\centering
  \includegraphics[width=0.9\linewidth]{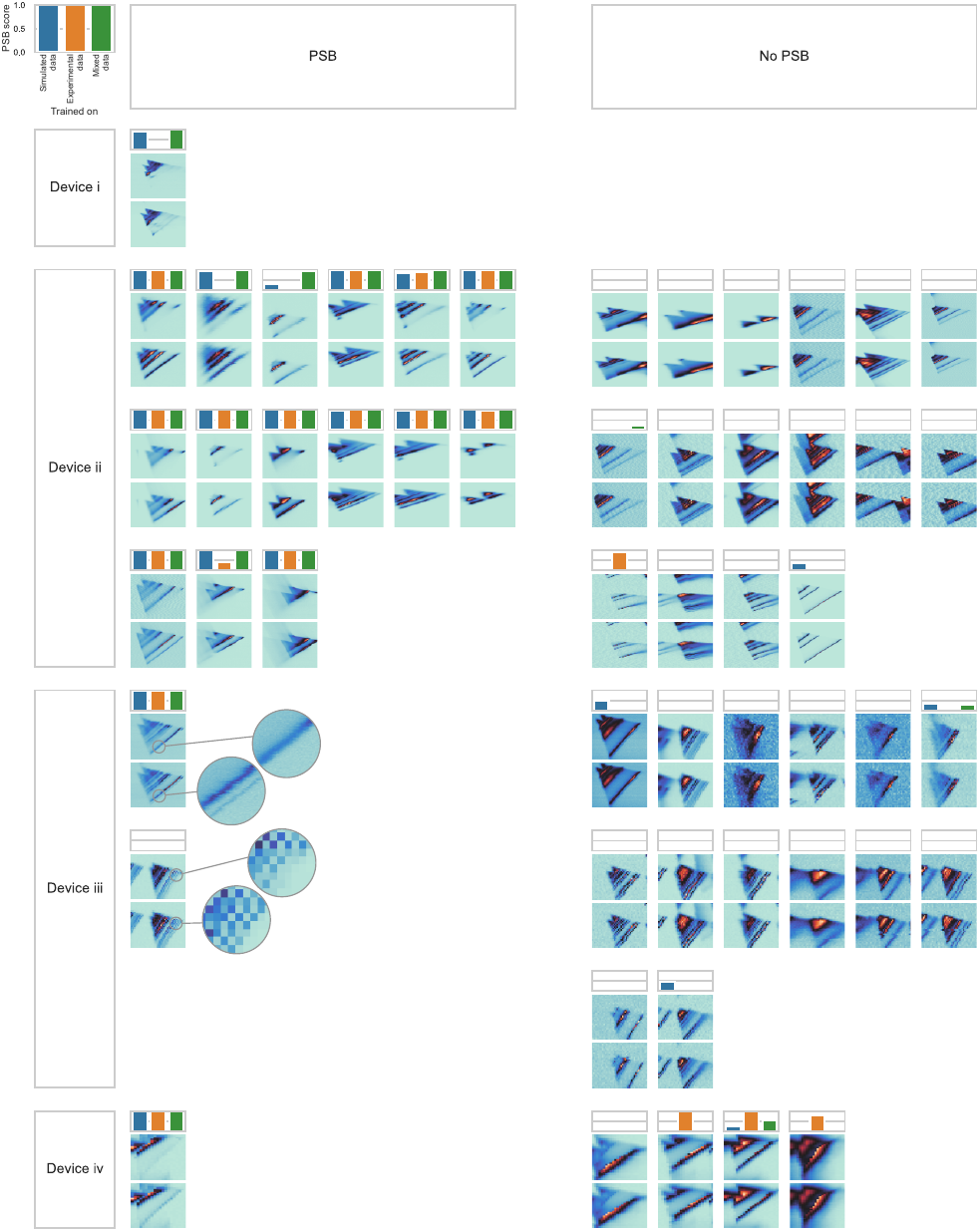}
  \caption{ \textbf{All experimental data used for benchmarking and the prediction of the ensembles of classifiers.} There are two main columns (sets with and sets without PSB) and four main rows (devices i to iv). The measurements shown in the left column (marked ``PSB'') are not definitive proof that PSB is present but show the signs we expect and want to detect with a classifier. In the top left corner there is an example prediction plot with all labels: We show ensemble scores 
  for the case of training on simulated data (blue), experimental data (orange), or a mix of both (green). The prediction threshold is plotted in the background as a horizontal line at a score of 0.5. Each set consists of a current measurement with low or zero magnetic field (top) and with a large magnetic field (bottom) and is jointly normalised between 0 and 1. Above each set the score plot is shown for that set without labels. The sets with PSB of device iii additionally show a magnification of a part of the base of the triangles due to visibility.}
  \label{fig:all_examples}
\end{figure*}

We show all experimental data used for the training and testing of the classifier in Fig.\,\ref{fig:all_examples}. We split them according to device source (row) and whether they show signs of PSB or not (column). Data from device iv corresponds to charge transitions A, B, C, D, and H with a low magnetic field of $B=-0.04\,$T (see Fig.\,\ref{fig:poc} and Fig.\,\ref{fig:more_clf_results}). We show three ensemble classifier scores (one for each type of training data) for each example. In the left column, we expect a perfect classifier to predict PSB (which would mean a full bar), in the right column it would predict no PSB (which would mean a missing bar).

This shows which examples were hard for the classifiers and it also shows the diversity of examples the classifier needed to deal with.

Some bias triangles show only weak signs of PSB, such as the ones from device iii. Even though we can not be sure from those measurements that this is indeed due to PSB and not due to other effects, e.g.~orbital effects, we still label those bias triangles as showing PSB. The point is that the signature could be due to PSB and should therefore be caught by a classifier. Confirmation measurements such as the ones done in Fig.~\ref{fig:poc}d need to be performed after that to verify the prediction.

\onecolumn

\end{document}